\documentclass[10pt,conference]{IEEEtran}
\usepackage[utf8]{inputenc}
\IEEEoverridecommandlockouts
\usepackage{cite}
\usepackage{multirow}
\usepackage{amsmath,amssymb,amsfonts}
\usepackage[edges]{forest}
\usepackage{algorithmic}
\usepackage{pgf-pie}
\usepackage{graphicx}
\usepackage{textcomp}
\usepackage{xcolor}
\usepackage{color, colortbl}
\usepackage{verbatim}
\usepackage{subcaption}
\usepackage{pgfplots, pgfplotstable}
\usepackage{url}
\usepackage{tikz}
\pgfplotsset{width=7cm,compat=1.17}

\def\BibTeX{{\rm B\kern-.05em{\sc i\kern-.025em b}\kern-.08em
    T\kern-.1667em\lower.7ex\hbox{E}\kern-.125emX}}

\begin{document}

\title{Exploring Relevant Artifacts of Release Notes: \\The Practitioners' Perspective}

\author{
\IEEEauthorblockN{ Sristy Sumana Nath}
\IEEEauthorblockA{\textit{Computer Science},
\textit{University of Saskatchewan}\\
Saskatoon, Saskatchewan, Canada  \\
sristy.sumana@usask.ca}
\and
\IEEEauthorblockN{ Banani Roy}
\IEEEauthorblockA{\textit{Computer Science},
\textit{University of Saskatchewan}\\
Saskatoon, Saskatchewan, Canada  \\
banani.roy@usask.ca}
}

\maketitle

\begin{abstract}
A software release note is one of the essential documents in the software development life cycle. The software release contains a set of information, e.g., bug fixes and security fixes.
Release notes are used in different phases, e.g., requirement engineering, software testing and release management.
Different types of practitioners (e.g., project managers and clients) get benefited from the release notes to understand the overview of the latest release. As a result, several studies have been done about release notes production and usage in practice.
However, two significant problems (e.g., duplication and inconsistency in release notes contents) exist in producing well-written \& well-structured release notes and organizing appropriate information regarding different targeted users' needs.
For that reason, practitioners face difficulties in writing and reading the release notes using existing tools.
To mitigate these problems, we execute two different studies in our paper. 
First, we execute an exploratory study by analyzing 3,347 release notes of 21 GitHub repositories to understand the documented contents of the release notes. As a result, we find relevant key artifacts, e.g., issues (29\%), pull-requests (32\%), commits (19\%), and common vulnerabilities and exposures (CVE) issues (6\%) in the release note contents.
Second, we conduct a survey study with 32 professionals to understand the key information that is included in release notes regarding users' roles. For example, project managers are more interested in learning about new features than less critical bug fixes.
Our study can guide future research directions to help practitioners produce the release notes with relevant content and improve the documentation quality. 

\end{abstract}

\section{Introduction}
Regular releases of software promise to improve product quality and high customer satisfaction \cite{Semiautomatic}.
Release notes inform users about essential project changes, e.g., bug fixes, feature enhancements, source code changes, from the previous version to the latest version \cite{arena1}.
Practitioners (e.g., project managers, developers and clients) use release notes in various software development phases, e.g., requirements engineering, software programming, debugging and testing phase \cite{Semiautomatic,softwaredocumentation}.
Several empirical studies have been done to analyze the release note contents \cite{rnempirical, releasenote, arena1}. For example, Moreno et al. \cite{arena1} identify 17 types of changes that can be included in the release notes, and Bi et al. \cite{rnempirical} classify those changes into eight categories.
Futhermore, Abebe et al. \cite{releasenote} discover additional six different types of information in release notes (described in Section \ref{background}). 
Despite the existing studies, in reality practitioners face difficulties in release note productions and usages because of \textit{unclear contents with poor structural presentations}. 
The documented information in release notes is currently
scattered and poorly organized, and the information is described vaguely \cite{rnempirical}. Consequently, release notes help limited release note users in many cases. Therefore, identifying relevant software artifacts and linking them with release notes can help to resolve this issue \cite{rnempirical}.  

The release management process consists of several activities from release initiating to closing \cite{End-UserPerspectivesinRel}. Different target users are involved in the release process, and
they are looking for different types of information regarding their project roles \cite{Semiautomatic} in release notes.
Bi et al. \cite{rnempirical} classify users into two categories: (i) {Release Note Producer} and (ii) {Release Note User}. Moreover, the authors cover significant discrepancies between release note producers and users in perceiving release notes. For example, release note producers focus on high-level changes and design decisions of development. In contrast, release note users expect detailed descriptions of new features compared to the previous release. 
Therefore, \textit{difficulties to use} is another problem of release notes usage. Because it is challenging to know the requirements for the release notes concerning the target users, conducting a survey study with practitioners can help understand the specific audiences' needs.

Our study has two motives: (1) to investigate the relevant software artifacts that can help to classify and structure the information in release notes; and (2) to understand the target users' requirements to tailor the release note contents. 
\begin{itemize}
    \item Exploratory Study: we extract and analyze 3,347 release notes of 21 GitHub projects and then separate release notes' contents (or sentences). We identify different artifacts related to the contents and classify the artifacts into three categories. The study design and result analysis are described in Section \ref{study1} and Section \ref{label:result1} respectively.
    \item Survey Study: we gather practitioners' opinions on release notes in practice and receive responses from 32 participants. We classify these participants into two categories: internal and external team members. The study design and result analysis are described in Section \ref{study2} and Section \ref{label:result2} respectively.
\end{itemize}

The key contributions of our study:
\begin{itemize}
\item We extract the data from GitHub and develop a dataset for the contents of release notes.
    \item We identify essential software artifacts from the dataset those help to produce well-structured release notes and classify the contents.
    \item We analyze the response of the participants and summarize them, which can aid tailoring release notes automatically for different stakeholders.
    
\end{itemize}



\section{ Background}\label{background}

\subsection{Release Note Contents}
Different communities produce release notes according to their own guidelines, and no common standards exist for writing release notes. In order to understand the release note contents (see Table \ref{tab:contentcategory}), several empirical studies have been done to investigate and categorize the documented information in release notes.
For example, Moreno et al. \cite{arena1} manually inspected 990 release notes to analyze and classify their content into 17 categories by focusing on information at finer level granularity.
Similarly, Bi et al. \cite{rnempirical} classified the documented information of release notes into eight main categories, comparatively higher-level classification.

On the other hand, Abebe et al. \cite{releasenote} identify six different information, e.g., titles, system overview, resource requirements, installation, addressed issues and caveat, that are included in release notes which contents are relatively high-level. Authors mainly focus on software end-users' perceptions of release notes. Klepper et al. \cite{Semiautomatic} identify some additional information, for example, technical information and testing instructions, in the release notes.
Our study explores the different software artifacts relevant to the release note contents.

\begin{table}[]
    \centering
    \scriptsize
    \caption{The categories of the documented information in release notes}
    \label{tab:contentcategory}
    \begin{tabular}{p{7em}|p{17em}|p{4em}}
    \hline
         \textbf{Study} & \textbf{Contents Category} &\textbf{Percentage} \\ \hline \hline
        \multirow{17}{*}{Moreno et al. \cite{arena1}} & Fixed Bugs & 90\%  \\ \cline{2-3}
&New Features &46\%\\ \cline{2-3}
&New Code Components &43\%\\ \cline{2-3}
&Modified Code Components &40\%\\ \cline{2-3}
&Modified Features&26\%\\ \cline{2-3}
&Refactoring Operations &21\%\\ \cline{2-3}
&Changes to Documentation &20\%\\ \cline{2-3}
&Upgraded Library Dep.&16\% \\ \cline{2-3}
&Deprecated Code Components &10\%\\ \cline{2-3}
&Deleted Code Components &9\%\\ \cline{2-3}
&Changes to Config. Files&8\%\\ \cline{2-3}
&Changes to Code Components Visibility &7\%\\ \cline{2-3}
&Changes to Test Suites&7\%\\ \cline{2-3}
&Known Issues&6\%\\ \cline{2-3}
&Replaced Code Components &5\% \\ \cline{2-3}
&Architectural Changes &3\%\\ \cline{2-3}
&Changes to Licenses&2\%\\ \hline
         \multirow{8}{*}{ Bi et al. \cite{rnempirical}} & Issues Fixed  & 79.3\%\\\cline{2-3}
         &New features & 55.1\%\\\cline{2-3}
        & System internal changes &25.1\%\\\cline{2-3}
        & Non-functional requirements &10.3\%\\\cline{2-3}
        & Documentation update &9.5\%\\\cline{2-3}
        & Configuration &2.8\%\\\cline{2-3}
        & Required further actions &2.1\%\\\cline{2-3}
        & Refactoring and reuse &1.9\%\\ \hline
    \end{tabular}
\end{table}


\subsection{Practitioners of Release Notes}

Different role-based practitioners are involved in a software release process \cite{Semiautomatic}. 
Bi et al. \cite{rnempirical} classify the stakeholders into two groups: release note producers and users. Architects and team managers are mainly responsible for producing release notes. 
Developers basically write release notes to reflect internal code changes, whereas testers and operators use release notes to perform their tasks.
The purpose of release notes usages for different stakeholders (e.g., developers and testers) regarding the phases are described below:
\begin{itemize}
    \item In the pre-alpha phase, \textit{project managers} and \textit{clients} use the release notes to discuss the project progress with new functionalities and significant issues.
   \item The release notes of an alpha version are used both by \textit{developers} and \textit{testers}. During this phase, developers first debug all the critical bugs found in a pre-alpha phase. Then, the testers assess the functionality of the application and compare the expected value with the final output value by utilizing the release notes.
   
   \item To test the system in the real environment, \textit{clients} and \textit{end-users} use the beta version of the release notes.
   
   \item \textit{Team managers} release a stable version of the system and highlight activities in the release notes of the RC version.
   
   \item Before deploying the final version of product, the team members write the release notes for the \textit{integrators} (who are using a library in their code) and \textit{end-users}. Therefore, the release notes of the final version need to be concise and properly understandable for the target audience.
\end{itemize}
However, the documented ineffective and unnecessary information creates  problems on the usage of release notes for the targeted audiences \cite{rnempirical}. Therefore, our study investigates the valuable information of release notes depending on the target users' perceptions.

\section{Study Design}\label{studydesign}
\begin{figure}
    \centering
    \includegraphics[width=3in]{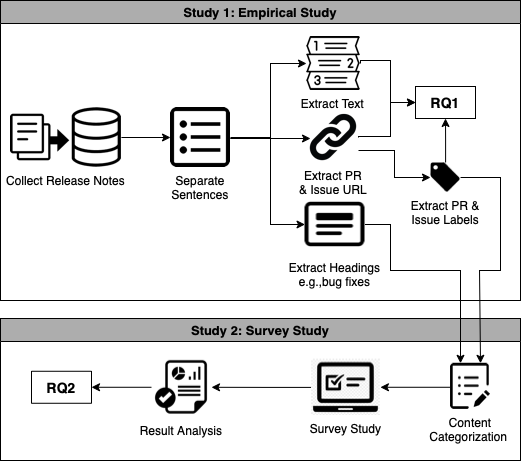}
    \caption{Overview of our study}
    \label{fig:studydesign}
\end{figure}

This section describes our research questions and study processes. Figure \ref{fig:studydesign} represents the overview of this study.

\subsection{Research Questions}
We address the following research questions:

\textbf{RQ1:} \textit{What software artifacts are important for preparing release notes?}
To answer the research question, we conduct an exploratory study on 3,347 release notes from 21 GitHub project repositories for understanding the release note contents.
Majority of the cases, we identify three crucial software artifacts, e.g., commits, issues and pull-requests, from GitHub and one artifact, e.g., common vulnerabilities and
exposures (CVE) issues, is extracted from external sources. These valuable sources can help to prepare quality release notes. Moreover, we find other key artifacts that can assist in maintaining well-structured release notes.

\textbf{RQ2:} \textit{What types of release note contents may vary depending on the software development role of practitioners?}
Everyone involved in a software development project can be a practitioner of release notes., and the addressed contents of the release notes can be modified depending on the users' particular information needs \cite{Semiautomatic}. To understand the targeted users' needs, we prepare an online survey study and learn users' opinions about release notes. This survey study may help produce high-quality release notes and tailor the documented release notes depending on users' needs.

\subsection{Study 1: Exploratory Study}\label{study1}
One key goal of our study is to identify the relevant artifacts of release notes (i.e., RQ1). Therefore, we targeted open-source projects and collected release notes from GitHub-hosted projects.
\subsubsection{Data Collection}
First, we collect release notes of open-source projects from GitHub. In this step, we use search option of GitHub by sorting the most number of stars of three-top languages, e.g., JavaScript, Java and Python, based on active repositories \cite{githut}. 
To eliminate the trivial projects, we define following criteria for a project selection from GitHub: 
\begin{itemize}
    \item \textbf{created:} minimum of three years ago and is active 
    \item \textbf{forks and stars:} $\geq$ 1,000 times forks and $\geq$  8,000 stars
    \item \textbf{contributors:} $\geq$ 30 committers
    \item \textbf{total commit:} $\geq$ 2,000 
    \item \textbf{release notes:} $\geq$ 30
    \item \textbf{total resolved issues:} $\geq$ 500 issues
    \item \textbf{total pull-requests:} $\geq$ 500 pull-requests
    
\end{itemize}

Then, we select total 21 projects, i.e., 8 JavaScript, 7 Java and 6 Python projects, based on the number of stars. Among the projects, we did not consider the non-engineering ones, e.g., \texttt{iluwatar/java-design-patterns} and \texttt{kdn251/interviews}.
Table \ref{tab:dataset} represents the detail information about the selected repositories. After project selection, we extract the release notes using the data extraction tool.
Data is available on GitHub \cite{dataavailable}.

\begin{table}[]
\caption{Dataset Overview}
    \label{tab:dataset}
    \centering
    \tiny
    \begin{tabular}{ p{1em} | p{11em} | p{15em}|p{5em}  }
    \hline
    \textbf{SL.} &\textbf{Repository} & \textbf{Domain}& \textbf{\# Releases} \\ \hline \hline
         1&vuejs/vue & Web libraries and frameworks & 210  \\
         \rowcolor{lightgray}
         2&facebook/react & Web libraries and frameworks & 96  \\
         3&twbs/bootstrap & Web libraries and frameworks & 73  \\
         \rowcolor{lightgray}
         4&axios/axios & Web libraries and frameworks & 32  \\
         5&nodejs/node & System software & 252  \\
         \rowcolor{lightgray}
         6&mrdoob/three.js & Web libraries and frameworks & 125  \\ 
         7&mui-org/material-ui & Web libraries and frameworks & 322\\
         \rowcolor{lightgray}
         8&chartjs/Chart.js & Web libraries and frameworks & 81  \\
         
         9&spring-projects/spring-boot & Software tools & 112  \\ 
         \rowcolor{lightgray}
         10&elastic/elasticsearch &Web libraries and frameworks & 60  \\
         11&ReactiveX/RxJava & System software & 225  \\
         \rowcolor{lightgray}
         12&google/guava & Non-web libraries and frameworks & 34  \\
         13&PhilJay/MPAndroidChart & Non-web libraries and frameworks & 44  \\
         \rowcolor{lightgray}
         14&redisson/redisson&Non-web libraries and frameworks & 103  \\
         15&jenkinsci/jenkins & System software & 128 \\
         \rowcolor{lightgray}
         16&tensorflow/tensorflow & Software tools & 152  \\
         17&tiangolo/fastapi  & Non-web libraries and frameworks & 113  \\ 
          \rowcolor{lightgray}
         18&getsentry/sentry  & Non-web libraries and frameworks &  38 \\
         19&pandas-dev/pandas  & Non-web libraries and frameworks & 81  \\
          \rowcolor{lightgray}
         20&apache/airflow & Software tools & 41  \\
         21&home-assistant/core & Application software & 847  \\\hline
         
    \end{tabular}
    
\end{table}

\begin{figure}
    \centering
    \includegraphics[width=3in]{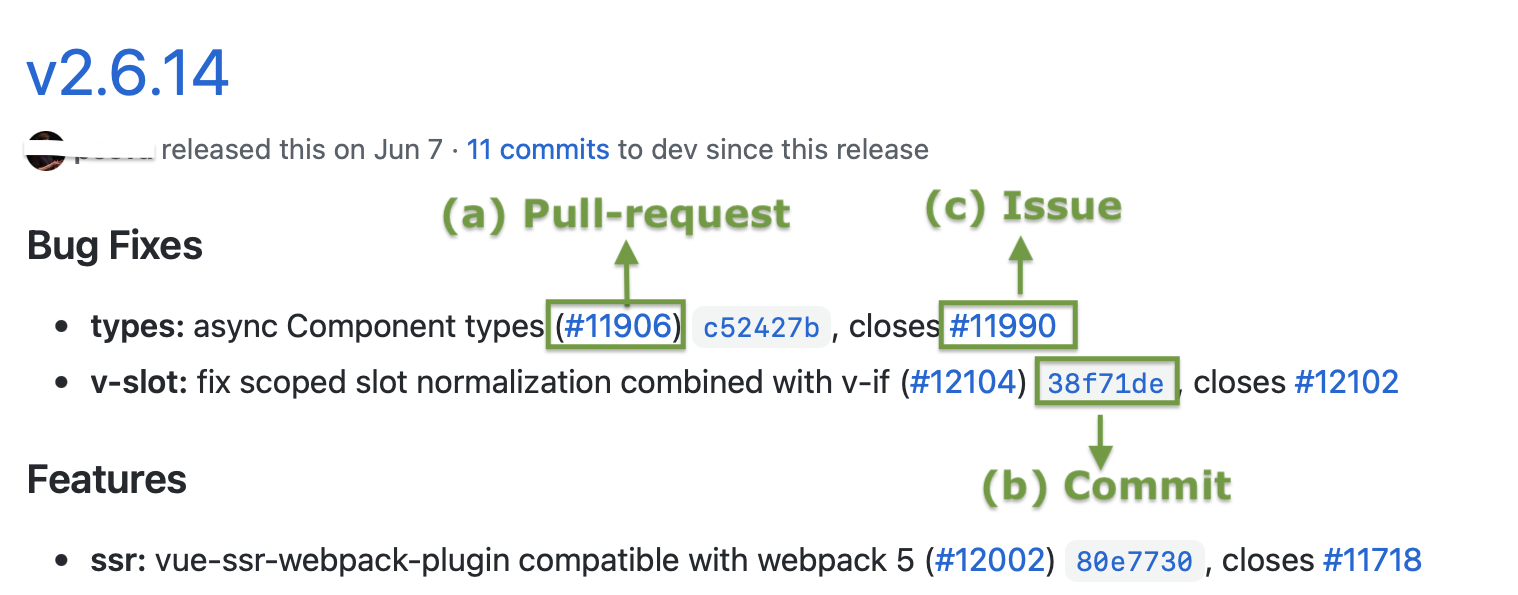}
    \caption{Artifacts of Release Notes}
    \label{fig:releasenotecontentartifacts}
\end{figure}

\subsubsection{Data Analysis}
First, we filter out the empty release notes from the extracted data. Second, we eliminate some information, e.g., contributors' name, to analyze the release note contents.
Third, we split the sentences, i.e., contents, and headings from release notes. For example, Fig. \ref{fig:releasenotecontentartifacts} represents a release note. Here, \textit{Bug Fixes, Features} are heading and the bullet listed information are contents. Then,
we extract the URLs from each sentences and maintain separate column in the dataset for further analysis.

\begin{figure*}
    \centering
    \begin{tikzpicture}
\begin{axis}[
    ybar stacked,
    width=7in,
    height=4in,
    bar width=15pt,
    ymin=0,
    nodes near coords,
    style={font=\footnotesize},
	nodes near coords style={ color=black, opacity=1},
    legend style={at={(0.5,-0.15)},
      anchor=north,legend columns=-1},
    ylabel={Contents},
    symbolic x coords={vue, react, bootstrap,axios, node, three.js, material-ui, chart.js, 
		spring-boot, elasticsearch, rxjava, guava, mpandriodchart, redisson, jenkins, tensorflow, airflow, fastapi, sentry, pandas, core},
    xtick=data,
    x tick label style={rotate=45,anchor=east},
    ]
\addplot+[black,fill opacity = .1]plot coordinates {(vue,517) (react, 68) (bootstrap, 30) (axios,0) (node,595)
  (three.js,568) (material-ui, 0) (chart.js,0) (spring-boot,0) (elasticsearch,0) (rxjava,0) (guava,150) (mpandriodchart,41) (redisson,0) (jenkins,45) (tensorflow,0) (airflow,0) (fastapi, 0)(sentry, 25)(pandas, 0)(core, 50)};
\addplot+[black,fill opacity = .27] plot coordinates {(vue,353) (react, 180) (bootstrap, 227) (axios,79) (node,393)
  (three.js,449) (material-ui, 48) (chart.js,300) (spring-boot,419) (elasticsearch,171) (rxjava,320) (guava,55) (mpandriodchart,0) (redisson,0) (jenkins,71) (tensorflow,0) (airflow,105) (fastapi, 215)(sentry, 52)(pandas, 0)(core, 178)};
\addplot+[black,fill opacity = .39] plot coordinates {(vue,205) (react, 0) (bootstrap, 596) (axios,23) (node,58)
  (three.js,90) (material-ui, 58) (chart.js,300) (spring-boot,639) (elasticsearch,589) (rxjava,70) (guava,93) (mpandriodchart,0) (redisson,10) (jenkins,55) (tensorflow,0) (airflow,0) (fastapi, 6)(sentry, 25)(pandas, 250)(core, 123)};
\addplot+[black,fill opacity = .45] plot coordinates {(vue,0) (react, 0) (bootstrap, 4) (axios,0) (node,163)
  (three.js,0) (material-ui, 0) (chart.js,0) (spring-boot,3) (elasticsearch,0) (rxjava,0) (guava,6) (mpandriodchart,0) (redisson,0) (jenkins,0) (tensorflow,405) (airflow,0) (fastapi, 0)(sentry, 0)(pandas, 0)(core, 3)};
  
\addplot+[black,fill opacity = .55] plot coordinates {(vue,0) (react, 7) (bootstrap, 0) (axios,32) (node,0)
  (three.js,0) (material-ui, 248) (chart.js,0) (spring-boot,0) (elasticsearch,9) (rxjava,0) (guava,104) (mpandriodchart,55) (redisson,530) (jenkins,18) (tensorflow,90) (airflow,5) (fastapi, 0)(sentry, 357)(pandas, 0)(core, 0)};
  
  \legend{\strut commits , \strut pull-requests , \strut issues , \strut CVE, \strut no links}
  \end{axis}
\end{tikzpicture}
    \caption{Release Note Contents Analysis}
    \label{fig:comparisoncontent}
\end{figure*}

\begin{table}[!hbt]
\centering
  \caption{Participants' Data}
  \label{tab:Participants}
  \begin{tabular}{c|c|c|l}
    \hline
    \textbf{Category}&\textbf{Sub-Category}&\textbf{Participants}&\textbf{Total}\\
    \hline \hline
    \multirow{2}{*}{\textbf{Type}}&Internal User&19&\multirow{2}{*}{32}\\
    &External User&13&\\
    \hline
    \multirow{3}{*}{\textbf{Experience}}&1-3 years&9&\multirow{3}{*}{32}\\
    &3-6 years&16&\\
    &More than 6 years&7&\\ 

  \hline
\end{tabular}
\end{table}
\subsection{Study 2: Survey Study}\label{study2}

The purpose of this survey study is to learn about effective information of the release notes based on users' opinion. In this survey, we recruit participants who have minimum one year software development experience. In particular, we choose the users who have knowledge in software release and release notes.
We adopt Non-Probabilistic Sampling methods, i.e., convenience sampling and snowball sampling, to invite participants of our study.
We followed two steps to invite participants:
{{(1)}} We sent email invitations to 200 contributors (only release note producers) of the selected GitHub projects in our study and received 11 (5.5\%) responses. In GitHub, one contributor writes release notes in the specific release (marked as green color in Fig. \ref{fig:releasenotecontentartifacts}).
{{(2)} }We invited professionals, e.g., project managers, testers, integrators, clients, within our social network and asked them to circulate this invitation to other professionals using sampling methods. We get 21 responses by following this strategy.

We prepare set of questions to professionals about their role and experience and their opinions about release notes.
We design a survey study using Google Form and asked participants about their roles and opinions in software development. Firstly, participants select
their specific role (e.g., project manager, client) and 
write the number of years of experience in software development process. Finally, they provide their responses of the questions related to release notes, e.g., the content specification and structure of release note and comment. 
We use the 4-point Likert Scale (based on \textit{Importance}) to understand the priority of the content of release notes based on a user role. Moreover, Abebe et al. \cite{releasenote} identify three different structures which are followed to prepare release notes, and we asked the participants to provide their feedback on the suitable structure of release notes. To response this question, participants select another 5-point Likert Scale (based on \textit{Quality}).
We have received responses from total 32 participants.
Our questionnaires can be found on GitHub \cite{dataavailable}.

\section{Result Analysis}\label{label:results}

We describe the results of the empirical study
(to answer to RQ1) in Section \ref{label:result1}. The results of professionals’ opinions on release notes are shown in Section \ref{label:result2} (to answer RQ2).

\subsection{RQ1: What software artifacts are important for preparing release notes?}\label{label:result1}
Several studies \cite{ARENA, arnnode.js, rnempirical, releasenote, releaseseke2021} analyze the release note production and usage; however, the exploration study about release note contents and identifying related software artifacts have not yet been investigated. 
To investigate the relevant artifacts, we analyze the contents of our developed dataset. 
We identify two types of information are important in the release notes. {One} is \textit{heading} of these contents, e.g., bug fixes and features. {Second} is \textit{contents}, i.e., list of resolved issues, e.g., \textit{v-slot: fix scoped slot normalization combined with v-if} (shown in Fig. \ref{fig:releasenotecontentartifacts}).
After pre-processing the text, our dataset contains 37,152 release note contents.
We mainly focus on this study to find the relevant software artifacts with contents; therefore, we preserve the heading for future work to classify the contents that can help maintain the well-structured information in the release notes. 

In the contents analysis step, we find 
some reference numbers along with URLs that are incorporated with contents (see in Fig. \ref{fig:releasenotecontentartifacts}).
We identify four different URLs pattern from the contents by extracting the URLs. Depending on these patterns, we detect four different software artifacts that can help to produce release notes.
For example, issues, pull-requests, commits and CVE (Common Vulnerabilities and Exposures) are linked with software release in Fig. \ref{fig:releasenotecontentartifacts}.  
For example, 
\begin{itemize}
    \item Issues: /issues/\texttt{issue\_id}
    \item Pull-requests: /pull/\texttt{pull\_id}
    \item Commits: /commit/\texttt{commit\_id}
    \item CVE: /cvename.cgi?name=\texttt{cve\_id}
\end{itemize}

Fig. \ref{fig:comparisoncontent} shows the comparison of these four type artifacts, and among them, the bar chart represents pull-requests (32\%) are the most important artifacts to produce release notes. 
Interestingly, in 41\% cases, the release note contents are linked with multiple artifacts, e.g., issues, pull-requests and commits. For example, the documented information of release notes is linked with multiple artifacts (see in Fig. \ref{fig:comparisoncontent}).
To improve the usability of release notes for practitioners, we need to consider the content selection approach from these artifacts in future.

\begin{figure}
    \centering
    \includegraphics[width=3in]{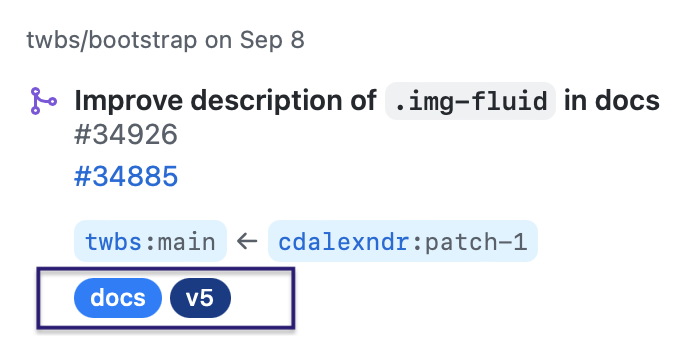}
    \caption{Labels}
    \label{fig:labels}
\end{figure}

Additionally, we explore the linked commits, pull-requests and issues and identify the labels of pull-requests and issues. For example, \texttt{docs} and \texttt{v5} are two labels of issues (see in Fig. \ref{fig:labels}). Cabot et al. \cite{labelsuse} analyze the issue labels and categorize them into four categories based on the usages, e.g., priority, versioning, workflow and architecture. Our analysis finds the text similarity between content headings and issues/pull-requests label. The contents come from the issues, pull-requests and commits (developers cannot add labels in commit); therefore, we can use the labels to classify the information in release notes. Table \ref{tab:labels} represents the example.

Finally, based on the analysis of result of RQ1, we classify these artifacts into three categories: 
\begin{itemize}
    \item \textbf{internal artifacts:} pull-requests, issue, commit
\item \textbf{external artifacts:} CVE issues  
\item \textbf{other key artifacts:} tags of issues/pull-request, milestones, change-log.md, wiki
\end{itemize}

Moreover, around 14\% cases, we do not find any related artifacts in the release note contents. In future, we need to analysis manually to get some more findings about release note contents.

\subsection{RQ2: What types of release note contents may vary depending on the software development role of practitioners?}\label{label:result2}

\begin{figure*}
    \centering
        \begin{tikzpicture}[scale=0.4]

 \pie [rotate = 180,color = {gray!30,gray!50,gray!10,gray!60}]
    {21/Project managers,
     53/Developers, 26/Testers}

\end{tikzpicture}
        \begin{tikzpicture}[scale=0.4]

 \pie [rotate = 180, 
 color = {gray!50,gray!10}]
    {61.5/Integrators,
     38.5/Clients}

\end{tikzpicture}
    \caption{Ratio of User's Participation}
    \label{fig:participation}

\end{figure*}
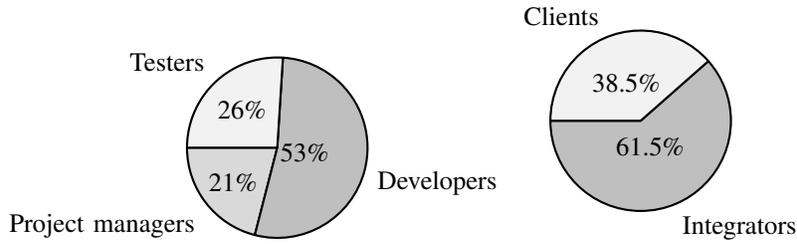
\begin{table*}[!hbt]
\begin{center}
\caption{Survey Result}
\scriptsize
\label{tab:result}
    \begin{tabular}{  p{13em} || p{4em} |p{2em} |p{4em} |p{2em} | p{4em} |p{2em} |p{4em} |p{2em} | p{4em} |p{2em}}
\hline
\multirow{2}{*}{\textbf{Types of Content}} & \multicolumn{10}{c}{\textbf{Practitioners' Response} (Likert Distributions \& Avg. Score)}\\ 
 & \multicolumn{2}{c|}{\textbf{Project Managers}} & \multicolumn{2}{c|}{\textbf{Developers}} & \multicolumn{2}{c|}{\textbf{Testers}} & \multicolumn{2}{c|}{\textbf{Integrators}} & \multicolumn{2}{c}{\textbf{Clients}} \\ \hline \hline
\textbf{Bug Fixes} & \includegraphics[width=0.05\textwidth]{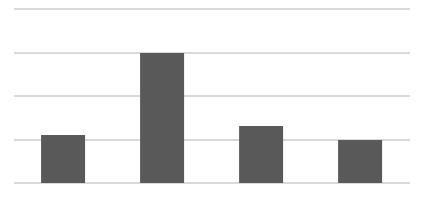}& 2.3 & \includegraphics[width=0.05\textwidth]{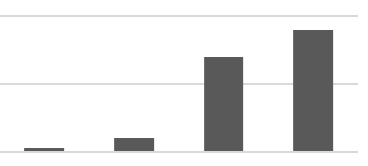} & 3.4 & \includegraphics[width=0.05\textwidth]{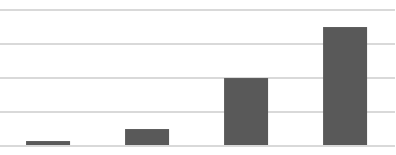} & 3.4 & \includegraphics[width=0.05\textwidth]{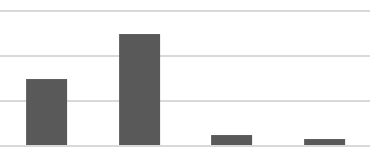} & 1.8& \includegraphics[width=0.05\textwidth]{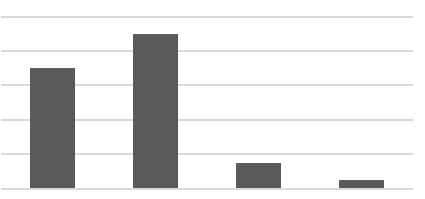} & 1.7 \\ \hline
\textbf{New Features} & \includegraphics[width=0.05\textwidth]{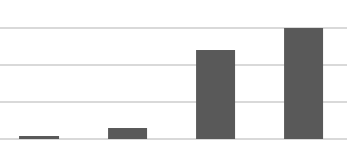}& 3.4 & \includegraphics[width=0.05\textwidth]{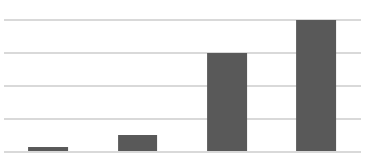} & 3.4 & \includegraphics[width=0.05\textwidth]{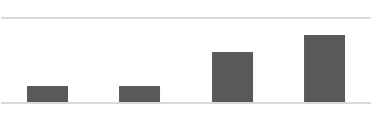} & 3.1 & \includegraphics[width=0.05\textwidth]{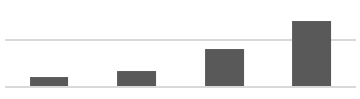} & 3.2& \includegraphics[width=0.05\textwidth]{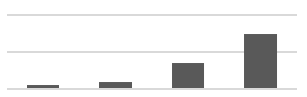} & 3.4 \\ \hline
\textbf{Feature Enhancements} & \includegraphics[width=0.05\textwidth,height=0.028\textwidth]{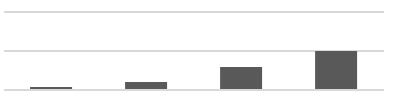}& 3.3 & \includegraphics[width=0.05\textwidth,height=0.028\textwidth]{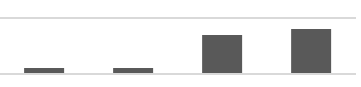} & 3.3& \includegraphics[width=0.05\textwidth,height=0.028\textwidth]{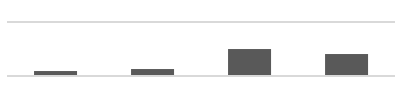} & 3.1 & \includegraphics[width=0.05\textwidth,height=0.028\textwidth]{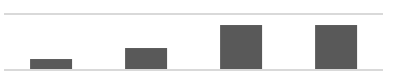} & 3.0& \includegraphics[width=0.05\textwidth,height=0.028\textwidth]{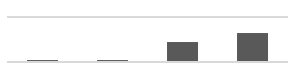} & 3.4 \\ \hline
\textbf{Documentation Changes} & \includegraphics[width=0.05\textwidth,height=0.03\textwidth]{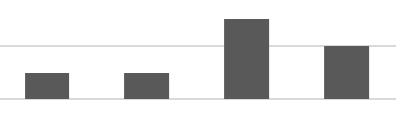}& 2.8& \includegraphics[width=0.05\textwidth,height=0.03\textwidth]{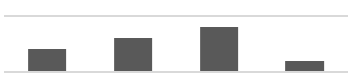} & 2.4 & \includegraphics[width=0.05\textwidth,height=0.03\textwidth]{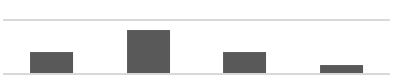} & 2.1 & \includegraphics[width=0.05\textwidth,height=0.03\textwidth]{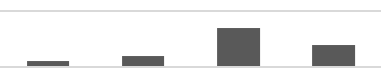} & 3.0& \includegraphics[width=0.05\textwidth,height=0.03\textwidth]{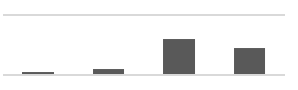} & 3.2 \\ \hline
\textbf{Internal Code Changes} & \includegraphics[width=0.05\textwidth,height=0.028\textwidth]{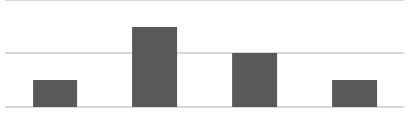}& 2.4 & \includegraphics[width=0.05\textwidth,height=0.028\textwidth]{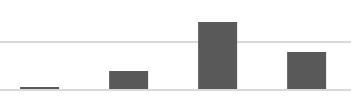} & 3.1 & \includegraphics[width=0.05\textwidth,height=0.028\textwidth]{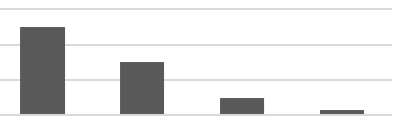} & 1.6 & \includegraphics[width=0.05\textwidth,height=0.028\textwidth]{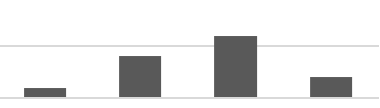} & 2.6& \includegraphics[width=0.05\textwidth,height=0.028\textwidth]{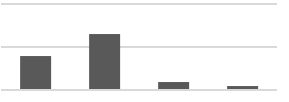} & 1.8 \\ \hline
\textbf{Architectural Changes} & \includegraphics[width=0.05\textwidth,height=0.028\textwidth]{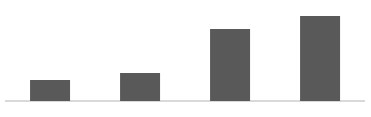}& 3.0 & \includegraphics[width=0.05\textwidth,height=0.028\textwidth]{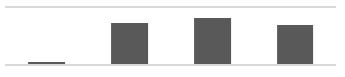} & 2.9 & \includegraphics[width=0.05\textwidth,height=0.028\textwidth]{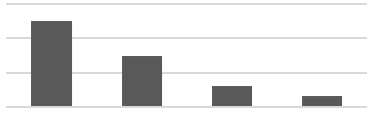} & 1.7 & \includegraphics[width=0.05\textwidth,height=0.028\textwidth]{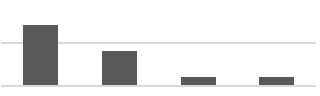} & 1.3& \includegraphics[width=0.05\textwidth,height=0.028\textwidth]{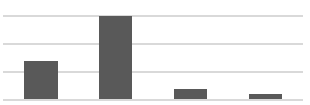} & 1.9 \\ \hline
\textbf{Security Issues} & \includegraphics[width=0.05\textwidth,height=0.028\textwidth]{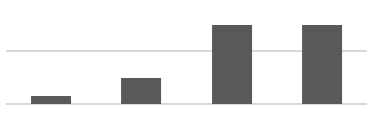}& 3.1 & \includegraphics[width=0.05\textwidth,height=0.028\textwidth]{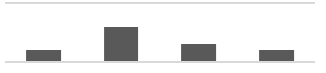} & 2.3 & \includegraphics[width=0.05\textwidth,height=0.028\textwidth]{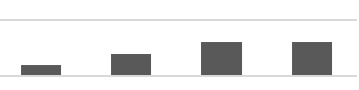} & 2.8 & \includegraphics[width=0.05\textwidth,height=0.028\textwidth]{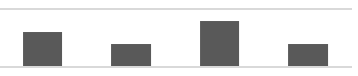} & 2.5& \includegraphics[width=0.05\textwidth,height=0.028\textwidth]{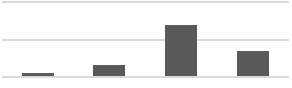} & 3.0 \\ \hline
\textbf{Performance Issues} & \includegraphics[width=0.05\textwidth,height=0.028\textwidth]{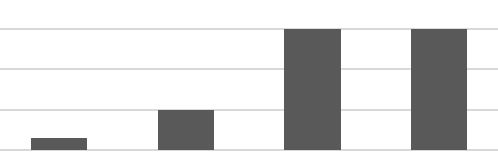}& 3.2 & \includegraphics[width=0.05\textwidth,height=0.028\textwidth]{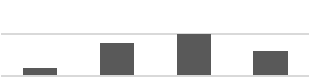} & 2.8 & \includegraphics[width=0.05\textwidth,height=0.028\textwidth]{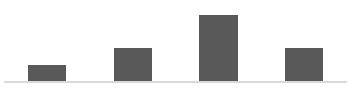} & 2.8 & \includegraphics[width=0.05\textwidth,height=0.028\textwidth]{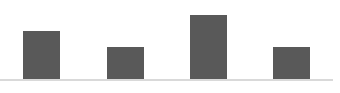} & 2.4& \includegraphics[width=0.05\textwidth,height=0.028\textwidth]{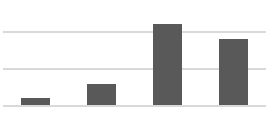} & 3.1 \\ \hline
\textbf{Dependency Changes} & \includegraphics[width=0.05\textwidth,height=0.028\textwidth]{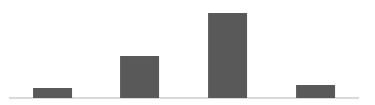}& 2.7 & \includegraphics[width=0.05\textwidth,height=0.028\textwidth]{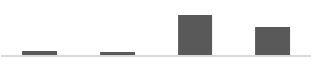} & 3.2 & \includegraphics[width=0.05\textwidth,height=0.028\textwidth]{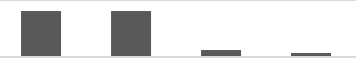} & 1.7 & \includegraphics[width=0.05\textwidth,height=0.028\textwidth]{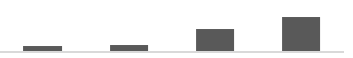} & 3.2& \includegraphics[width=0.05\textwidth,height=0.028\textwidth]{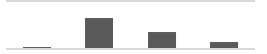} & 2.5 \\ \hline
\textbf{Configuration Changes} & \includegraphics[width=0.05\textwidth,height=0.028\textwidth]{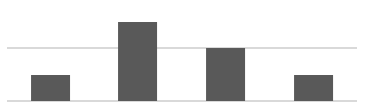}& 2.4 & \includegraphics[width=0.05\textwidth,height=0.028\textwidth]{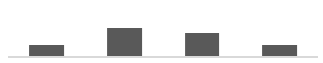} & 2.4 & \includegraphics[width=0.05\textwidth,height=0.028\textwidth]{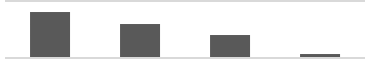} & 1.8 & \includegraphics[width=0.05\textwidth,height=0.028\textwidth]{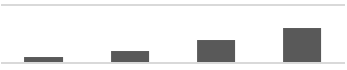} & 3.1& \includegraphics[width=0.05\textwidth,height=0.028\textwidth]{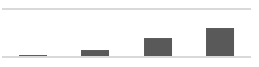} & 3.3 \\ \hline
\textbf{Tests} & 
\includegraphics[width=0.05\textwidth,height=0.028\textwidth]{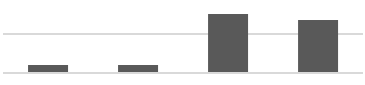}& 3.2 & \includegraphics[width=0.05\textwidth,height=0.028\textwidth]{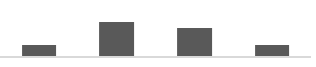} & 2.4 & \includegraphics[width=0.05\textwidth,height=0.028\textwidth]{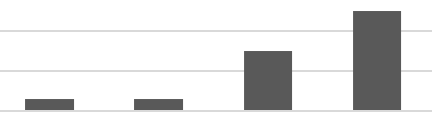} & 3.3 & \includegraphics[width=0.05\textwidth,height=0.028\textwidth]{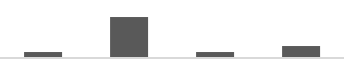} & 2.4& \includegraphics[width=0.05\textwidth,height=0.028\textwidth]{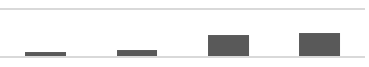} & 3.1 \\ \hline


\end{tabular}
\end{center}
\end{table*}

We conduct a survey study with professionals and define a set of questions about release notes contents and structure. In this study, we need expert opinions, and for that reason, we encourage those participants who  have at least one year experience in software development.
From the previous studies \cite{arena1, rnempirical} (details in Table \ref{tab:contentcategory}), we can see that bug fixes and new features are documented more in the release notes than the other types of contents. 
Our survey study includes these content types and asks participants what contents categories they are looking for along with the justifications.
Table \ref{tab:result} presents the content categories and summary of survey results.

According to their opinion, software release notes can be customized that will help for documenting for target users. Participants choose option of 4-Likert Scale (\textit{not important, slightly important, moderately important, important}).
We received the responses of 32 participants of five different role-based users. 
The results of the online survey are summarised in Table \ref{tab:result}. We have found the following observations: 
\begin{itemize}
    \item Bug fixes are essential content for \textit{developers} and \textit{testers} based on the avg. score (i.e., 3.4). Besides, \textit{Project managers} and \textit{clients} mention they are interested to know learn major or critical issues from the release notes rather than the all bugs. However, list of all bugs is not helpful. 
    \item New features and feature enhancements are important content of release notes for both internal users and external users (Avg. score of likert distributions of Table \ref{tab:result}).
    \item \textit{Integrators} and \textit{clients} concern about the documentation changes, e.g., installation instructions and UI changes, in the software development.
    \item The internal code changes include added/modified /deleted classes/methods in the source code. Developers must document these changes in release notes because it will help track changes, and new developers can benefit. Moreover, integrators want to know the source code changes that can help them integrate into their system.
    \item Architecture provides fundamental structure of software development and development teams follow the architectural process, e.g., model-view-controller model, to build the software. Refactoring is one kind of architectural changes; this process is restructuring the code without changing the external behavior. \textit{Project managers} and \textit{developers} give priority to know about architectural changes through the release notes.
    \item \textit{Project managers} and \textit{clients} want to know the non-functional changes, e.g., security issues or performance improvements, about the software. Because these changes help to improve the quality of software by solving some vulnerabilities and optimization issues.
    \item Dependency changes mean upgrade the APIs or libraries. This content essential for \textit{developers} and \textit{integrators} because without updating the dependency changes they cannot implement or use the updated version of the software.
    \item \textit{Integrators} and \textit{clients} want to learn about the required configuration changes (e.g., installation or hardware requirements) because without updating these requirements they cannot use the latest release.
    \item Testing information, e.g., test cases or suites, is useful content of release notes for \textit{testers}. Because the internal testers receive unstable versions and can know which tasks to be tested or which to be ignored.
\end{itemize}

\begin{figure}
    \centering
  \begin{tikzpicture}
\begin{axis}[
ybar,
height=5cm,
	x tick label style={
		/pgf/number format/1000 sep=},
	ylabel=\#number of response,
	bar width=7pt,
enlarge x limits=0.20,
x tick label style={rotate=20,anchor=east},
	legend style={
    at={(0.5,-0.23)},
    anchor=north,legend columns=-1
},
ymin=0,
ybar,
xtick=data,
symbolic x coords={very poor, poor, fair, Good, Excellent},
grid=major,
xmajorgrids=false
]
\addplot [black,fill,fill opacity = .6] 
	coordinates {(very poor,2) (poor,1) (fair,9)
		  (Good,14) (Excellent,6)};
\addplot [black,fill,fill opacity = .4] 
	coordinates {(very poor,4) (poor,3) (fair,11)
		  (Good,7) (Excellent,7)};
\addplot [black,fill,fill opacity = .2] 
	coordinates {(very poor,9) (poor,10) (fair,5)
		  (Good,2) (Excellent,3)};
\legend{Style 1, Style 2, Style 3}
\end{axis}
\end{tikzpicture}

  \caption{Response about the Structure of Release Notes} 
  \label{fig:style} 
\end{figure}
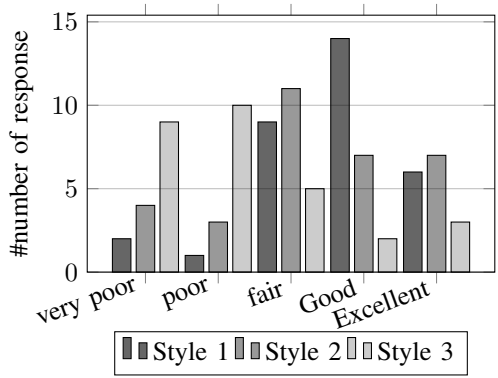

 Additionally, we ask the participants about the suitable structure of release notes, which is easy to read. The following three styles are observed in \cite{releasenote}:
\begin{itemize}
    \item Style 1: summarize or re-phrase the addressed issues in the current release
    \item Style 2: list of selected issues that were addressed in the current release (do not summarize issues)
    \item Style 3: list of all addressed issues in the current release
\end{itemize}
Participants provide score according to the quality structure (\textit{very poor, poor, fair, good, excellent}) of release notes. We identified style 1, e.g., summary of related issues is more preferable (i.e., 14 and 6 users choose Good and Excellent option, respectively) than the other two structures (see Fig. \ref{fig:style}). Example of statements:
\begin{itemize}
    \item \textit{``Need to use few sentences with bullet points to let users know what’s new and how to navigate these."}
    \item \textit{``Major issues of the release are required to include in release notes; however, sometimes technical contexts are made hard to read the release notes."}
\end{itemize}

\section{Discussions}\label{label:discussion}
We discuss our results analysis findings and threat to validity in this section.
\subsection{Release Notes Generation}
We discuss few concerns about release notes generation.

\textbf{Content Selection.}
The results of RQ1 show that release note contents are linked with issues, pull requests, and commits. Generally, issues are created first, and developers solve those issues through commits and pull requests. Release notes of the latest version contain the resolved issues list from the immediate previous release. However, we identify that the documented contents can be different among issue titles, pull-request titles, and commit messages.
We represent one content from the release note of version 2.6 \cite{vue2.6.13} of Vue GitHub project. 
\begin{itemize}
\item \textbf{Content:} \textit{attrs: do not consider translate attribute as boolean}
\item \textbf{Commit message:} \textit{attrs: do not consider translate attribute as boolean} \cite{vuecommit}
\item \textbf{Issue title:} \textit{isBooleanAttr for the HTML attribute 'translate'} \cite{vueissue11391}
\item \textbf{Pull-request title:} \textit{fix \#11391: translate attribute had incorrectly the key as it's value} \cite{vuepull11392}
\end{itemize}
The above example shows that the documented content comes from the commit messages, and the titles of issue and pull-request are different. Content selection is a crucial part of producing good-quality release notes. Different automated release notes generation tools \cite{ARENA, arnnode.js} consider only issues and generate text from source code changes. We encourage researchers could consider the content selection from the relevant artifacts before developing new tools.

         

\begin{table}[]
    \centering
    \caption{Examples of Common Content Types}
    \label{tab:labels}
    \scriptsize
    \begin{tabular}{p{15em}|p{15em}}
    \hline
          \textbf{Examples of headings} & \textbf{Examples of labels}\\ \hline \hline
         
          Fixed, Bug fixes & bug, regression, type:bug, bug on dependency library
         \\\hline
          Breaking changes, Features & feat:ssr, feature request, type:feat
         \\\hline
          Development, Added, Modified & core, type:dev, type:chore\\\hline
          Docs, Documentation changes &  docs, type:documentation
         \\\hline
          Performance Improvements & performance, type:performance
         \\\hline
         Dependency Changes, API Changes, Upgrades, Dependencies & dependencies, type:dependency-upgrades
         \\\hline
    \end{tabular}
\end{table}

\textbf{Content Classification \& Structure.}
Bi et al. \cite{rnempirical} encourage researchers to identify other software artifacts which help to classify and structure the information in release notes. We identify that the issues and pull-requests labels can help to classify the release note contents. Table \ref{tab:labels} represents the similarity between contents' headings and tags. Labels or tags can help to classify the release note contents. 
Moreover, we identify another two artifacts, e.g., milestones and wiki, from \textit{Three.js}, \textit{redisson} and \textit{jenkins} projects. 
Similarly, setting a milestone of upcoming releases can organize the structure of documenting information of release notes. 

\textbf{Content Tailoring.}
There are no worldwide standards or guidelines for documenting release notes \cite{arena1,rnempirical}. Furthermore, documenting all issues or changes in release notes makes it too lengthy. Consequently, readers lose interest in knowing the critical addressed issues in the current release \cite{releasenote}.
Our second study finds that different practitioners look for different information in the release notes. Therefore, automated techniques need to integrate tailoring tools to customize the release note contents regarding target users' requirements. 

\subsection{Issues of Existing Release Notes}
\begin{table}[]
    \centering
    \caption{Problems of Existing Release Notes}
    \label{tab:challenges}
    \begin{tabular}{c|c|c|c}
    \hline
         \textbf{Language} &\textbf{Num. of Projects} & \textbf{Duplication} (\%) & \textbf{Inconsistency} (\%)  \\\hline
         JavaScript&8&19 &10\\
         Java&7&22 &7\\
         Python&6&0 &0\\
         \hline
    \end{tabular}
    
\end{table}
We identify two problems in the existing release notes in GitHub and that are discussed as follows.

\begin{figure}
    \centering
    \includegraphics[width=2.6in, height=2in]{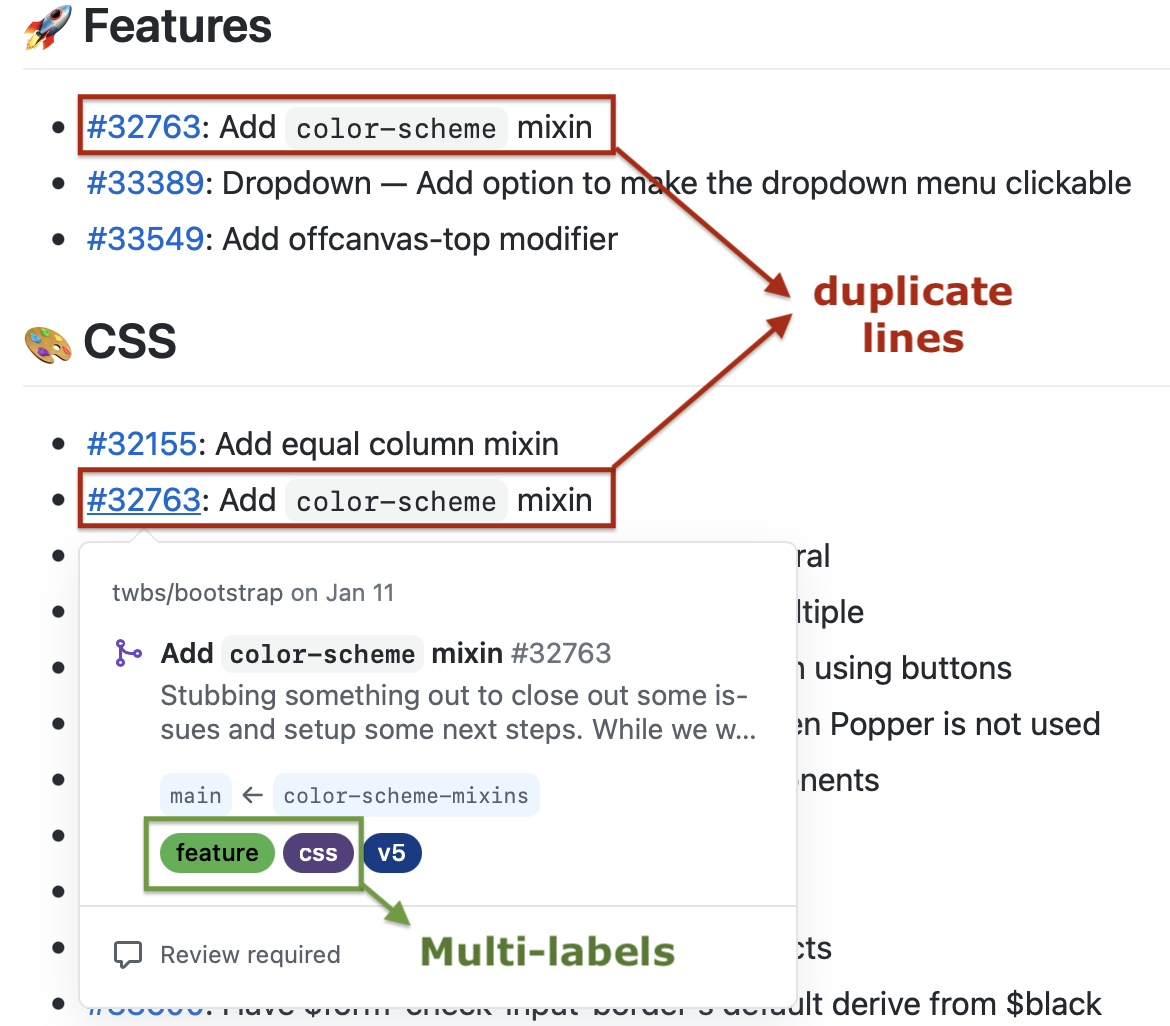}
    \caption{Content Duplication}
    \label{fig:duplicateline}
\end{figure}

\begin{figure}
    \centering
    \includegraphics[width=2.7in]{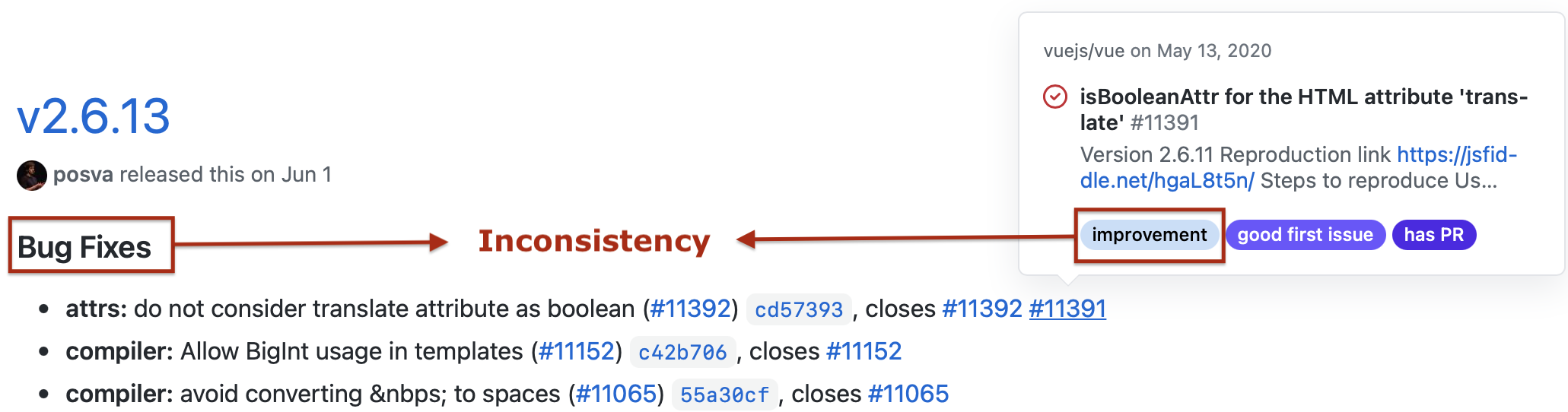}
    \caption{Inconsistency}
    \label{fig:inconsistency}
\end{figure}

\textbf{Content Duplication.} The first issue is same contents (or sentences) appear double or triple times in a release note. 
Fig. \ref{fig:duplicateline} represents an example of duplication problem and Table \ref{tab:challenges} shows the duplication percentage in different GitHub projects.
We investigate the reason behind this issue, and identify that issues are tagged by multiple labels, e.g., feature and css. 
Surprisingly, we notice that the duplication issue has not appeared in Python projects' release notes. 
Therefore, we analyze the release notes of the python projects and find that release note producers prepare release note contents manually. For example,
a {\textit{content}} from the release note of version 1.3.0 of Pandas is \textit{Performance improvement in DataFrame.corr() for method=kendall} and this content link with issue. However, the \textit{issue title} is \textit{DataFrame.corr(method="kendall") calculation is slow}
However, producing release notes manually takes lengthy time around 2 to 4 hours \cite{ARENA}. In future, we need to consider the duplication problems to prepare the release notes using automated tools.

\textbf{Inconsistency between Content Headings and Labels.} The second problem is an inconsistency between the contents' headlines and tags. 
Our analysis identifies the similarity between the contents' headlines and tags. Table \ref{tab:labels} shows the examples of headings and issue tags. We can see that those issues are labelled as \textit{bug, type:bug}, usually those are listed under \textit{Bug Fixes} or \textit{Fixed} headings. However, in some cases, we find inconsistency. Fig. \ref{fig:inconsistency} represents an example of inconsistency issue. For example, an issue is labelled by \textit{improvement} tag; however, this issue is written under \textit{Bug Fixes} heading in the release note. 
It is possible to eliminate the inconsistency issue, while the automated release notes generation techniques will be considered the labels to classify the contents.


\subsection{Threats to Validity}\label{threat}
\textbf{Internal validity-} The content of release notes can be different based on the project domains, e.g., application software and software tools.
Our study selects popular open-source projects' repositories in GitHub based on the most stars (i.e., more than 8,000 stars) and these projects belong to different domains. We analyze the three top most popular languages' projects, e.g., JavaScript, Java and Python, (details in Table \ref{tab:dataset}) based on the active repositories to mitigate the internal threat. Moreover, we sort the projects based on the total number of releases and mitigate those with less than 30 releases.

\textbf{External validity-} The total number of participants of this survey study is may not be enough to understand the real-life scenarios about release notes usage in software development practice. To mitigate this threat, on average, our study participants have 3.5 years of experience in software development and use version control systems, e.g., Git, to keep track of the changes in the system.

\section{Related Work}\label{label:relatedwork}
\textbf{Empirical Studies.}
Many empirical studies have been done in prior works. Among them, some studies investigate the documented information types in release notes \cite{rnempirical, releasenote, arena1} (detailed discussion in Section \ref{background}). 
Moreover, they identified that the provided information could differ in major and minor release notes from system to system.
Klepper et al. \cite{releasenote} propose a technique for manually tailoring release notes by a release manager while considering three different kinds of stakeholders, e.g., users, customers and team members in agile software development. 

\textbf{Tools and Techniques.}
Several studies aim to generate release notes automatically or semi-automatically \cite{ARENA, arnnode.js, releaseseke2021, Semiautomatic}.
A semi-automatic approach proposed by Klepper et al. \cite{Semiautomatic} for generating audience-specific release notes. This approach used issue tracker, version control system and build server as data sources for release notes generation and the release manager tailored the contents for different target users.
Moreno et al. \cite{ARENA} proposed another tool (i.e., ARENA) for Java projects. 
This tool uses external tools, e.g., change extractor, issue extractor, commits-issues linker and code change summarizer, to generate the complete release notes.
Ali et al. \cite{arnnode.js} follow a similar approach to generate release notes for Node.js projects. Both are language-specific tools.
Contrary, Nath et al. \cite{releaseseke2021} proposed a release notes generation technique by applying text summarization techniques (i.e., TextRank) using related commits and pull-requests. This approach selects top-ranked contents. Therefore, some important sentences might be missed to generate automated release notes. From prior studies, we find that the proper investigation is required to understand the input sources for the documented contents of existing release notes. 
Therefore, we investigate more than 3,000 release note contents and structure from GitHub.

\section{Conclusion}\label{conclusion}
We conduct an exploratory study and a user survey study on the release note contents to improve the quality of release notes production based on different practitioners' needs because, which are not explored in the existing studies.
Our first study identifies several artifacts, e.g., commits, issues and pull-requests, from our empirical study which can be linked with release note contents. In addition, we detect other key artifacts, e.g., tags and milestones, that can assist in classifying and maintaining an exemplary structure of the contents.
Since different practitioners get benefits by using release notes, it is essential to understand what information needs to be included for different users.
Keeping this in mind, we conduct a survey study to understand users' specific requirements.
Moreover, we discuss our findings of release note production and existing issues of the release note contents.
We are currently working on the three additional research questions to
extend our research:
(i) \textit{What is the efficient way to generate 
release notes automatically using the relevant artifacts?}; (ii) \textit{How can we integrate user-specific requirements to produce release notes?}; and (iii) \textit{How to assess the quality of release notes?}

\section*{Acknowledgment}
This research is supported by the Natural Sciences and Engineering Research
Council of Canada (NSERC), and by two Canada First Research Excellence
Fund (CFREF) grants coordinated by the Global Institute for Food Security
(GIFS) and the Global Institute for Water Security (GIWS).

\bibliographystyle{plain}
\bibliography{sample-base}
\end{document}